\documentclass[a4paper,12pt,twocolumn]{article}
\usepackage{cite}
\usepackage[dvips,hiresbb]{graphicx}
\usepackage{here}
\renewcommand{\figurename}{Fig.}

\title{\textbf{Special reflecting features of a three-dimensional photonic crystal depending on cut surfaces oriented in the direction [001] within photonic band gap}}
\author{\Large{Chikara Sakurai} \thanks{Email: c-sakurai@river-ele.co.jp (first), sakuraikazan@gmail.com (second)} \\ \textit{River Electric Corporation, 2-1-11 Fujimigaoka, Nirasaki, Yamanashi, Japan}} 
\date{\today}
\begin{document}
\onecolumn{
\maketitle
\begin{abstract}
\normalsize{The reflecting features (reflected spectrum and phase) on the surface of pure-material plate depend on refractive index, extinction coefficient, incident angle, polarization direction and so on. The features, however, do not depend on a cut surface that is another uniformly processed surface in the same direction, which is obtained by being cut, etched and polished for example. In this work, the special reflecting features of a three-dimensional (3D) terahertz photonic crystal (PC) depending on the cut-surface oriented in the same direction [001], have been studied especially within photonic band gap by using a FEM (finite element method). The 3D-PC is a silicon inverse diamond structure whose lattice point shape is vacant regular octahedrons. The characteristic features were studied on four types of cut-surface that were shifted per a/4 (a: lattice constant) in the same direction [001], especially with two types of orientation, [(1,1) and (1,$\overline{1}$) in an XY-plane] of electric-field of the incident wave in normal incidence. 
It was found that eight types of reflected spectrum and those of phase in all four cut-surface types are able to be classified in two patterns, they are correlated with two types of unit cell suggested, and the reflecting features invert or do not invert according to the cut-surface.} 
\end{abstract}
\twocolumn
\section*{Introduction}
\hspace*{5mm}The reflecting features (reflected spectrum and phase) on the surface of pure-material plane-parallel-plate with no inclusions and defects, depend on refractive index, extinction coefficient, incident angle, polarization direction and so on. The features, however, do not depend on a cut-surface that is another uniformly processed surface (mirror surface) obtained by being cut, etched and polished in the same direction.\\
\hspace*{5mm} Taking silicon (Si) for an example, the material belongs to cubic system, and
it is optically isotropic. The reflectivity is $(n-1)^2/(n+1)^2$, where n is refractive index, in the case of normal incidence and no absorption. The formula does not depend on the cut-surface.\\
\hspace*{5mm}In this work, special reflecting features of a three-dimensional (3D) terahertz photonic crystal (PC) depending on the \textit{cut-surface} oriented in the same direction [001], have been studied by using a FEM (finite element method). The 3D-PC is a silicon inverse diamond structure whose lattice point shape is vacant regular octahedrons.\\
\hspace*{5mm}Another special reflecting features of the 3D-PC have been studied in previous works, ref.~\cite{sakurai1} to \cite{sakurai4} \footnote{ Ref.~\cite{sakurai3} contains more detailed analyses than those in ref.~\cite{sakurai4}.}. These features on the polarization anisotropy are related to the polarization orientation (electric-field direction) difference between  reflected wave and incident one.\\
\hspace*{5mm}Measured polarization anisotropy$^{[3], [4]}$ in normal incidence indicated that the polarization orientation of a reflected wave rotated by 90 degrees to that of an incident wave at a frequency within BGX that is X point's photonic-band-gap (PBG).\\
\begin{figure}[t]
\centering
\includegraphics[width=7.8cm,]{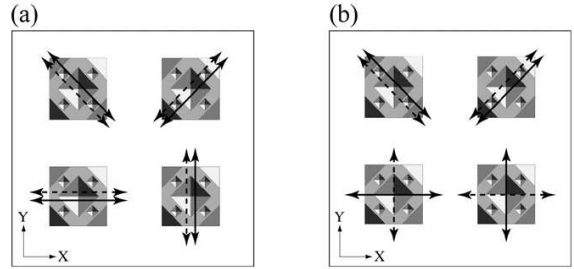}
\caption{\footnotesize (a) General polarization features within PBG expected from basic and optical rules. (b) Special polarization characteristic features within PBG. The incident wave is normal incidence, [001]. The heavy solid arrow is the polarization orientation (electric-field direction) of the incident wave, and the heavy dashed arrow is that of the reflected wave. The unit cell of a diamond structure in fig.~\ref{fig:latticeA} is illustrated.}
\label{fig:twopolar}
\end{figure}
\hspace*{5mm}Fig.~\ref{fig:twopolar}$^{[1]}$ shows the comparison of the polarization characteristic features as generally expected with ones in the measured spectra, within PBG.
Four types of polarization orientation of the incident wave are illustrated in each figure.\\
\hspace*{5mm}None the less because X- and Y-directions, in other expression, (10)- and (01)-directions, normal  each other on the plane (001) are constitutionally and optically identical \textit{in appearance}, the polarization orientation of the reflected wave rotated by 90 degrees from that of the incident wave at a frequency within BGX when that of the incident wave was X- or Y-direction.\\
\hspace*{5mm}In ref.~\cite{sakurai1}, the phase-difference, size and orientation of electric-field of two reflected waves whose polarization orientations are (11)-direction and (1$\overline{1}$)-direction, which are also identical \textit{in appearance}, have been studied primarily and quantitatively for resolution of polarization rotation above.\\
\hspace*{5mm}In addition, it was found that above phase-difference gradually shifts when the frequency shifts within BGX, and then the reflected wave approximately becomes near-circularly-polarized wave at another frequency within BGX.\\
\hspace*{5mm}Polarization features of the 3D woodpile PC in ref.~\cite{min} also have been studied in the limit of low-frequency wave and without PBG. The 3D-PC has eigen modes (branches) without PBG. An eigen mode have an intrinsic symmetry, and some polarization anisotropy also exists.\\
\hspace*{5mm}The existence of two-dimensional modes (surface modes) within PBG in the 3D woodpile PC has been studied in ref~\cite{ken}. Whether a surface state contributes to special characteristic features of the reflected spectra in this and previous works or not is not known.
The reflectivity, however, was not 100$\%$ within BGX, and twenty to forty percentage decrement was found though the transmittance was nearly equal to zero within the FEM analysis.\\
\hspace*{5mm}In ref.~\cite{pri}, the direction- and wavelength-dependent polarization anisotropy of stop gap branching has been studied with fcc symmetry as a function of an incidence angle for two polarization orientation normal each other, TE and TM. The electric field of TE (TM)  is oriented perpendicular (parallel) to the plane of incidence. The phase and polarization rotation of the reflected wave are not mentioned.\\
\hspace*{5mm}In this work, the reflected spectra and these phases in four types of cut-surface and four types of orientation of electric-field of the incident wave in fig.~\ref{fig:twopolar}, have been studied with the FEM. They were correlated with two types of unit cell suggested. It was found that the reflected spectra and these phases are able to be classified in two patterns, and they invert or do not invert according to the cut-surface.\\
\section*{Parameters}
\begin{figure}[t]
\centering
\includegraphics[width=6.5cm,]{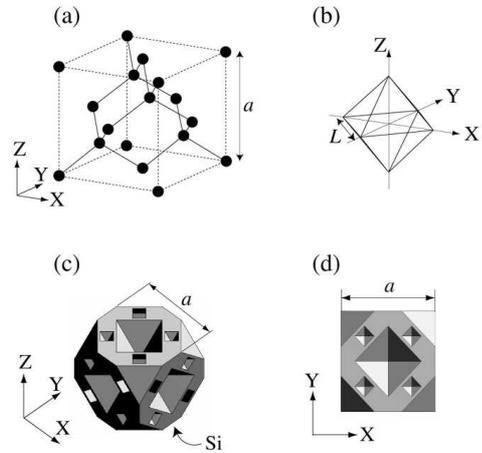}
\caption{\footnotesize (a) Lattice of the diamond structure. (b) Shape of the lattice point is regular octahedrons. It is vacant. The surrounding material is Si. (c), (d) Unit cell of the Si inverse diamond structure.}
\label{fig:latticeA}
\end{figure}
\hspace*{5mm}The lattice of the diamond structure in the simulation model, which is simply called unit cell, is shown in fig.~\ref{fig:latticeA}(a)\footnote{Parameters in this chapter are all the same as those in ref.~\cite{sakurai1} except cut-surface illustration in fig.~\ref{fig:cutsurface}}. The sphere is the lattice point and its shape is the regular octahedrons in fig.~\ref{fig:latticeA}(b). 
It is vacant (atmosphere) and the dielectric constant, $\varepsilon_1$ = 1.00 is set. 
The surrounding material is pure Si and the dielectric constant, $\varepsilon_2$ = 11.9 is set.
The lattice constant, $a$ = 300 $\mu$m and the side length of the regular octahedrons, $L$ = 150 $\mu$m are set.
In the unit cell of the Si inverse diamond structure, many sections of the air octahedrons on the lattice points are arranged as shown in fig.~\ref{fig:latticeA}(c) and \ref{fig:latticeA}(d). \\    
\begin{figure}[t]
\centering
\includegraphics[width=7.0cm, trim=0 0 0 0]{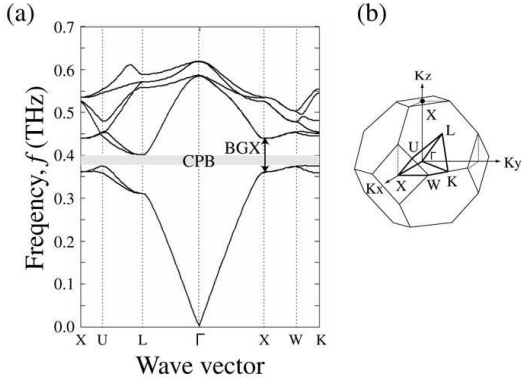}
\caption{\footnotesize (a) Calculated photonic band structure has complete photonic band gap, CPB (gray zone) at around 0.4 THz. BGX exists between 0.36 THz and 0.44 THz. (b) First Brillouin zone and the reduced zone (heavy line) with high symmetry points.}
\label{fig:bandstructure}
\end{figure}
\begin{figure}[t]
\centering
\includegraphics[width=4cm, trim=0 0 0 0]{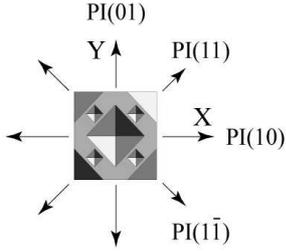}
\caption{\footnotesize Four types of PI(XY) is the polarization orientation of the incident wave in normal incidence.}
\label{fig:IXY}
\end{figure}
\hspace*{5mm}Fig.~\ref{fig:bandstructure}(a) shows the calculated photonic band structure~$^{[3]}$ by using plane wave expansion method.
The direction of the incident wave is +Z-direction, [001] in the real space and it corresponds to $\Gamma$-X direction in the wave number space (K-space). 
The reflected spectra and phases were studied on the surface (001)\footnote{[001] and (001) are defined on the X-Y-Z coordinate system in fig.~\ref{fig:latticeA}.} at around BGX. \\
\hspace*{5mm}Fig.~\ref{fig:IXY} shows the definition of polarization orientation (electric-field direction) of the incident wave, PI(XY). The unit cell of the diamond structure in fig.~\ref{fig:latticeA} is illustrated.
For example, PI(1, 0) and PI($-$1, 0) are indistinguishable in measurement; they are collectively expressed as PI(10). The direction of PI(01), PI(11) and PI(1$\overline{1}$) is (01)-, (11)-, and (1$\overline{1}$)-direction, respectively.\\
\begin{figure}[t]
\centering
\includegraphics[width=7.5cm, trim=0 0 0 -15]{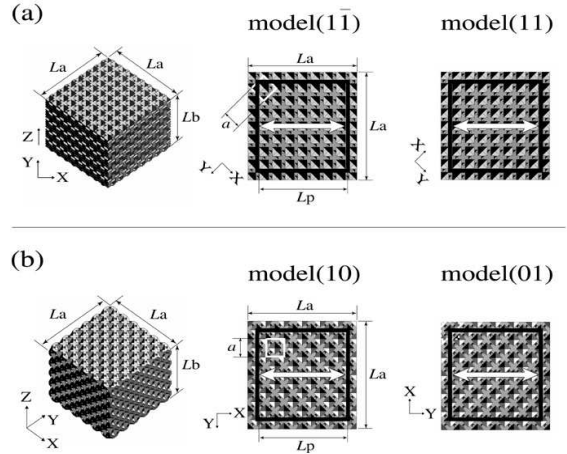}
\caption{\footnotesize In four analyzed models, the square surrounded with black heavy lines, port is the area of the incident wave. Each white heavy arrow is the orientation of electric-field vector of the incident wave, which is fixed. Instead, the models are rotated. }
\label{fig:modelAB}
\end{figure}
\begin{figure}[t]
\centering
\includegraphics[width=6.8cm, trim=0 0 0 -15]{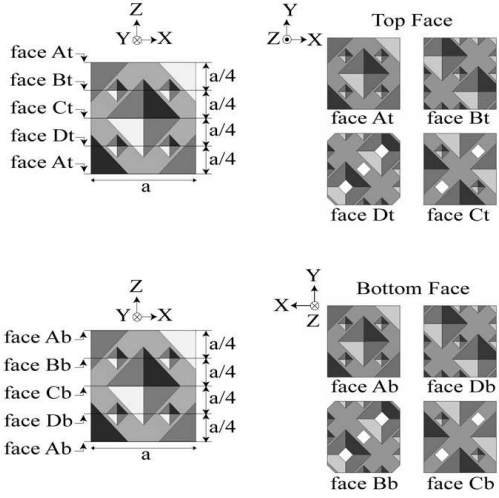}
\caption{\footnotesize Illustration of top face and bottom face of four types of cut-surface. Four types are type A to type D. For example, top face and bottom face of type A are face At and face Ab, respectively. The height is 5a, and the area in X-Y plane is 6a $\times$ 6a, which are common to four types.}
\label{fig:cutsurface}
\end{figure}
\hspace*{5mm}Four analyzed models are shown in fig.~\ref{fig:modelAB}. X-Y-Z coordinate system corresponds to that in fig.~\ref{fig:latticeA}.
The model size is $L_\mathrm{a}$ $\times$ $L_\mathrm{a}$ $\times$ $L_\mathrm{b}$ ($\mu$m)$^3$. $L_\mathrm{b}$ is the height of the Z-direction. $L_\mathrm{a}$ $\times$
$L_\mathrm{a}$ is the area in X-Y plane. 
The square area surrounded with white lines is $a$ $\times$ $a$ ($\mu$m)$^2$, which corresponds to the unit cell in fig.~\ref{fig:latticeA}(d).
The area of the incident wave, port is shown as the square surrounded with black heavy lines, whose size is $L_\mathrm{p}$ $\times$ $L_\mathrm{p}$ ($\mu$m)$^2$. $(L_\mathrm{a}, L_\mathrm{b}, L_\mathrm{p})$ is set as $(6a, 5a, 5a) (\mu$m), which is fixed. Models' mesh size is 50 $\mu$m\footnote{The mesh size is nearly minimum for~the limit of the personal computer's performance: OS: Windows~10~Home (64bit), processor: Intel(R) Core (TM) i5-6500 CPU@3.20 GHz, RAM: 64.0 GHz.}. Each white heavy arrows is the orientation of electric-field vector of the incident wave, which is fixed.\\
\hspace*{5mm}The direction of the incident wave is +Z-direction and normal incidence. Each model is rotated instead of the electric-field vector of the incident wave.
In model(11), model(1$\overline{1}$), model(10), and model(01), the orientation of electric-field of the incident wave is PI(11), PI(1$\overline{1}$), PI(10), and PI(01), respectively.\\
\hspace*{5mm}Fig.~\ref{fig:cutsurface} shows the explanation figures of  the cut surfaces common to each model in fig.~\ref{fig:modelAB}.
In this work, the reflecting features in four types (type A to type D) having different cut-surface are studied in each model in fig.~\ref{fig:modelAB}.\\
\hspace*{5mm} That the length, $(L_\mathrm{a}, L_\mathrm{b}, L_\mathrm{p})$ is  $(6a, 5a, 5a) (\mu$m) is common to all the types and models. Top face and bottom face of type A are face At and face Ab, respectively. Type A is original type, which has already been analyzed in ref.~\cite{sakurai1}. 
Those of type B are face Bt  and face Bb, which are shifted by a/4 in the Z-direction from 
those of type A, respectively. 
Similarly, those of type C (type D) are face Ct and face Cb (face Dt and face Db), which are  shifted by 2a/4 (3a/4). 
\section*{Simulation Results}
\hspace*{5mm}In this work, the reflecting features were primarily analyzed by using the FEM \footnote{The FEM software was Femtet(R)2017.1 made in Murata Software Co., Ltd..}. The transmission spectra were nearly equal to zero within BGX.
The FEM software had no command to obtain S-polarization (S-p) spectra and P-polarization (P-p) ones, separately. Only compound spectra of S-p and P-p were able to be obtained.
As definition, S-p is the orientation of electric-field of the reflected wave parallel to that of the incident wave which is shown as the white heavy allow in fig.~\ref{fig:modelAB}.  The orientation of P-p is perpendicular to that of S-p. \\
\hspace*{5mm}The reflecting features in figs.~\ref{fig:typeAB11} (a) to \ref{fig:typeAB11} (d) are obtained by using model (11) and model (1$\overline{1}$) in fig.~\ref{fig:modelAB}(a). Two types of reflectivity, R and those phases, $\theta$ (closed circle and open circle) and two types of transmittance, T (solid line and dashed line) are shown. The variable, $f$ is THz-frequency.\\
\hspace*{5mm}The analyzed frequency range is 0.25 THz to 0.55 THz, in which BGX (0.36 THz to 0.44 THz) is included. The frequency interval of the data is 0.005 THz. The absorption coefficient (cm$^-$$^1$) is set as zero. Two vertical solid lines are ones at 0.374 THz and 0.414 THz, which are explained in fig.~\ref{fig:typeABd}.\\
\hspace*{5mm}The symbol (XY) of the reflectivity, R(XY) and transmittance, T(XY), corresponds to that of PI(XY) in fig.~\ref{fig:IXY}.
$\theta$(XY) is the phase of R(XY), which is the phase-difference between the incident wave and reflected one, in model(XY) in fig.~\ref{fig:modelAB}.\\
\hspace*{5mm}In figs.~\ref{fig:typeAB11} (a) and \ref{fig:typeAB11} (c), PBG's are nearly equal to BGX (0.36 THz to 0.44 THz). The transmittance in each model is very small within PBG.   
In the sub-peaks of reflectivity and transmittance without PBG, local maximum value of the former and local minimum one of the latter tend to appear at the same frequency as expected.\\
\hspace*{5mm}In the shape of reflectivity of Type A within PBG, R(1$\overline{1}$) displays convex-downward one, and R(11) does concavity and convexity in fig.~\ref{fig:typeAB11} (a).\\
\hspace*{5mm}The phases, $\theta$(11) and $\theta$(1$\overline{1}$) of type A in fig.~\ref{fig:typeAB11} (b)
have different phase-frequency features especially within PBG. $\theta$(11) is larger than $\theta$(1$\overline{1}$), and two phases within PBG show a gradual decline with frequency compared with ones without PBG. A frequency at which the phase-difference between $\theta$(11) and $\theta$(1$\overline{1}$) is 180 degrees or 90 degrees, exists.\\
\begin{figure}[t]
\centering
\includegraphics[width=7.4cm, trim=10 0 0 0]{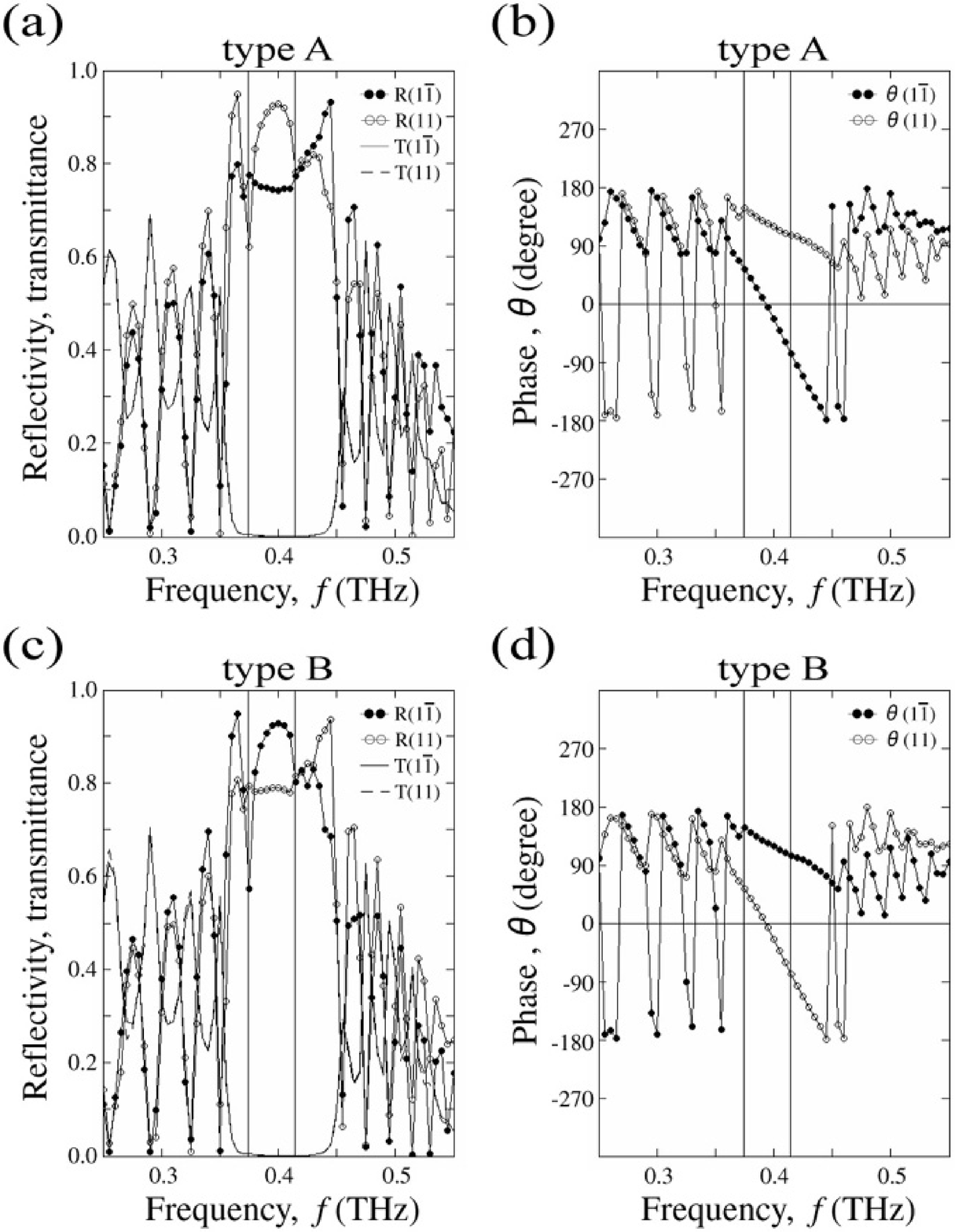}
\caption{\footnotesize Reflected spectra, R, those phases, $\theta$, and transmission spectra, T of type A and type B in model(1$\overline{1}$) and model (11).}
\label{fig:typeAB11}
\end{figure}
\hspace*{5mm}On the other hand, the special reflecting features are found by comparing the reflected spectra and phases of type A with those of type B.  Each reflected spectrum, R(11) and R(1$\overline{1}$) of type A and type B reverse each other. The phases, $\theta$(11) and $\theta$(1$\overline{1}$) also reverse. In other words, R(11), R(1$\overline{1}$), $\theta$(11) and $\theta$(1$\overline{1}$) of type B is nearly equal to R(1$\overline{1}$), R(11),  $\theta$(1$\overline{1}$) and $\theta$(11) of type A, respectively.\\
\begin{figure}[t]
\centering
\includegraphics[width=7.4cm, trim=10 0 0 0]{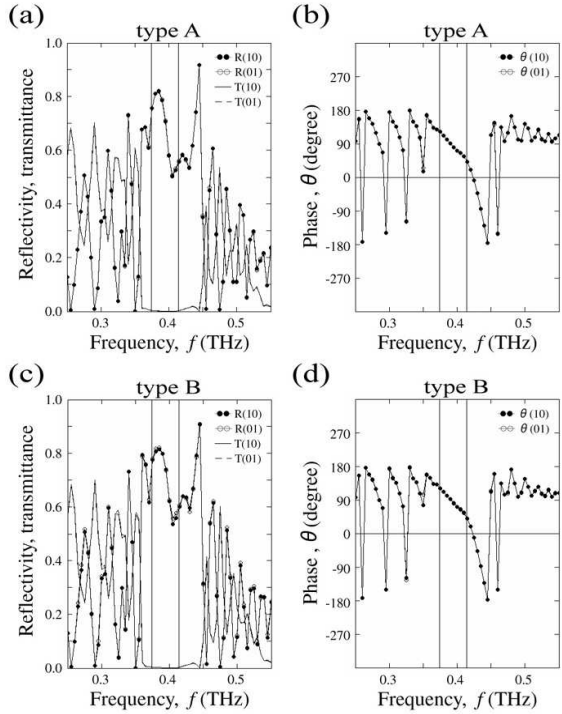}
\caption{\footnotesize Reflected spectra, R, those phases, $\theta$, and transmission spectra, T of type A and type B in model(10) and model (01).}
\label{fig:typeAB10}
\end{figure}
\begin{figure}[t]
\centering
\includegraphics[width=7.4cm, trim=10 0 0 5]{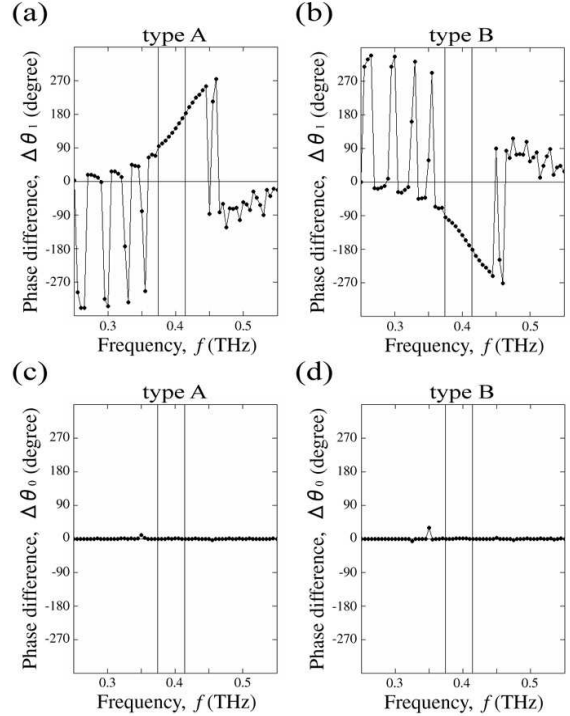}
\caption{\footnotesize Phase-differences of type A and type B. (a), (b) $\Delta\theta_{1}$ is the phase-difference, $\theta(11)-\theta(1\overline{1})$.  (c), (d) $\Delta\theta_{0}$ is $\theta(10)-\theta(01)$.}
\label{fig:typeABd}
\end{figure}
\hspace*{5mm}Fig.~\ref{fig:typeAB10} shows the reflecting features obtained by using model (10), model (01) in fig.~\ref{fig:modelAB}(b). The reflected spectra, R(10) and R(01) of type A are nearly equal each other, and these phases are also equal. In addition, the reflected spectra and these phases of type A and type B have little difference.\\ 
\hspace*{5mm}Fig.~\ref{fig:typeABd}(a) and \ref{fig:typeABd}(b) shows the phase-difference of type A and type B, respectively. The phase-difference, $\theta(11)-\theta(1\overline{1})$ is shown as $\Delta\theta_{1}$. In fig.~\ref{fig:typeABd}(c) and \ref{fig:typeABd}(d), the phase-difference, $\theta(10)-\theta(01)$ of type A and type B is shown as $\Delta\theta_{0}$.\\ \hspace*{5mm}The phase-difference, $\Delta\theta_{1}$ within PBG  gradually shifts compared with $\Delta\theta_{1}$ without PBG.  The size (absolute value) of $\Delta\theta_{1}$ is nearly equal to 90 degrees and 180 degrees at 0.374 THz and at 0.414 THz, respectively. Plus and minus of $\Delta\theta_{1}$ of type A and type B within PBG reverse. On the other hand, $\Delta\theta_{0}$ is almost constant and zero.
Two vertical solid lines at 0.374 THz and 0.414 THz are ones where the size of the phase-difference is nearly equal to 90 degrees and 180 degrees, respectively.\\
\hspace*{5mm}The reflecting features of type C and type D which correspond to fig.~\ref{fig:typeAB11} to  fig.~\ref{fig:typeABd} are shown in fig.A.~\ref{fig:typeCD11} to  fig.A.~\ref{fig:typeCDd} in appendix.
\section*{Discussions}
\hspace*{5mm} When the orientation of electric-field of the incident wave is PI(10) or PI(01), the orbits of electric-field of the reflected-wave in type A and type B are able to be calculated  by using basic and simple analysis with the phase-difference, $\Delta\theta_{1}$ and the reflectivity, R(11) and R(1$\overline{1}$); the orientation of electric-field of the reflected wave is \textit{parallel} to that of the incident wave when
that of the incident wave is PI(11) or PI(1$\overline{1}$), which is shown in fig.A.~\ref{fig:EEfield}\footnote{They are analyzed in detail in ref~\cite{sakurai1}. The electric-field vectors, which are illustrated as cones, are compound electric-field vectors of the incident and reflected waves in the FEM software used.}.\\
\hspace*{5mm}Fig.~\ref{fig:typeABorbit} shows the orbit of electric-field vector of the reflected wave at 0.374 THz and 0.414 THz in type A and type B. In type A, the orbit of electric-field of the reflected wave is near-circular ellipsoidal orbit at 0.374 THz, and that is linear-polarization orbit at 0.414 THz. They are shown as heavy lines in fig.~\ref{fig:typeABorbit} (a) and (b). The orbits of type B are shown as thin lines.\\
\hspace*{5mm}The phase-difference, $\Delta\theta_{1}$ of type A and that of type B are inverse relation each other. It means that rotational directions of electric-field of the reflected waves in two types are reverse each other. 
The orbit of type C (type D) is almost the same as that of type A (type B), which is shown in fig.A.~\ref{fig:typeCDorbit}.\\
\begin{figure}[t]
\centering
\includegraphics[width=7.0cm, trim=0  0 0 0]{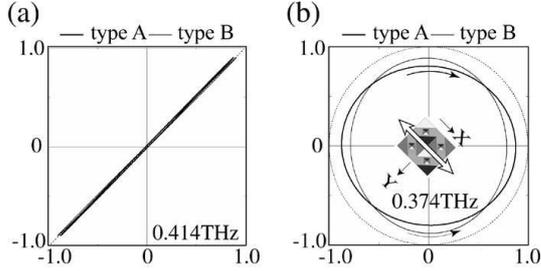}
\caption{\footnotesize Orbit of electric-field vector of the reflected wave in type A and type B. The white heavy arrow is the orientation of electric-field vector of the incident wave.
(a) at 0.414THz. (b)  at 0.374THz. The dashed line is true circle.}
\label{fig:typeABorbit}
\end{figure}
\begin{figure}[t]
\centering
\includegraphics[width=7.0cm, trim=0  0 0 0]{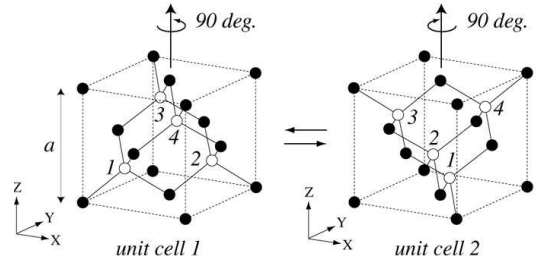}
\caption{\footnotesize Two types of unit cell. Unit cell 1 and unit cell 2 convert each other by 90-degree rotation about the central axis parallel to the Z-axis.}
\label{fig:unitcell12}
\end{figure}
\begin{figure}[t]
\centering
\includegraphics[width=6cm, trim=0 0 0 -50]{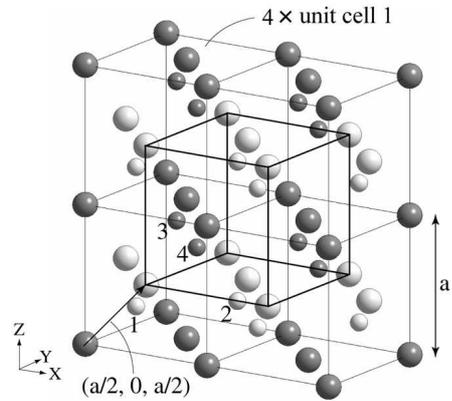}
\caption{\footnotesize Four unit cell 1 and  the unit cell of type C surrounded by heavy-line cubic frame. The latter unit cell is also unit cell 1.}
\label{fig:2x2x1unitcell}
\end{figure}
\hspace*{5mm}Fig.~\ref{fig:unitcell12} shows two types of unit cell. Unit cell 1 corresponds to the unit cell in fig.~\ref{fig:latticeA}.  The unit cell of type A belongs to it. Unit cell 2 is obtained by rotating unit cell 1 by 90 degrees about the central axis parallel to the Z-axis.
The X-Y-Z coordinate system is not rotated. \\
\hspace*{5mm}The method for preparing the unit cell of type C is shown in fig.~\ref{fig:2x2x1unitcell}.  Four fine-line cubic frames and  one heavy-line cubic frame and lattice points are illustrated.\\ 
\hspace*{5mm}Large dark spheres correspond to lattice points located on face At (face Ab).  Large white spheres correspond to ones located on face Ct (face Cb). Similarly, small dark spheres and small white ones correspond to lattice points located on face Bt (face Bb) and face Dt (face Db), respectively.\\
\hspace*{5mm}The group of the lattice points surrounded by the fine-line cubic frame corresponds to unit cell 1. The heavy-line cubic frame is obtained by shifting one unit cell 1 by (a/2, 0, a/2). 
Large dark and white spheres, and small dark and white ones are all comparable lattice points.
From the illustration, the unit cell of type C also belongs to unit cell 1.\\
\hspace*{5mm}By using the similar method, the unit cell of type B and that of type D belong to unit cell 2.
For example, the method for preparing the latter is shown in fig.~\ref{fig:2unitcell}.\\
\hspace*{5mm}From the view point of classification by two types of unit cell, each reflecting feature (reflected spectrum and phase) of type A and type C conforms, and that of type B and type D also does.\\
\hspace*{5mm}Furthermore, R(11), R(1$\overline{1}$), $\theta$(11) and $\theta$(1$\overline{1}$) of type A
(type C) is equal to R(1$\overline{1}$), R(11),  $\theta$(1$\overline{1}$) and $\theta$(11) of type B (type D), respectively. The reasons are as follows.\\
\hspace*{5mm}The orientation of electric-field of incident wave of the reflected spectrum, R(11) in type A is PI(11). That of R(1$\overline{1}$) in type A is PI(1$\overline{1}$). R(11) and R(1$\overline{1}$) in type A is spectra reflected from unit cell 1.
The alternative view is that R(1$\overline{1}$) in type A is correlated with the spectrum reflected from  unit cell 2 when the orientation of the incident wave is PI(11) since PI(11) and PI(1$\overline{1}$) are normal each other.\\
\begin{figure}[t]
\centering
\includegraphics[width=6cm, trim=0 0 0 -50]{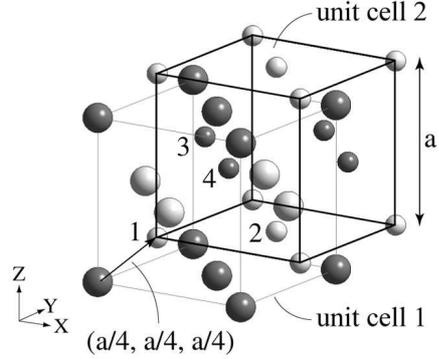}
\caption{\footnotesize One unit cell 1 and  the unit cell of type D surrounded by heavy-line cubic frame. The latter is unit cell 2.}
\label{fig:2unitcell}
\end{figure}
\hspace*{5mm}Similarly, when the orientation of the incident wave is fixed as PI(11), R(11)
and R(1$\overline{1}$) of type B are correlated with the spectrum reflected from  unit cell 2 and unit cell 1, respectively.\\
\hspace*{5mm}Accordingly, from the same view point, R(11) of type A (type C) conforms R(1$\overline{1}$) of type B (type D), and R(1$\overline{1}$) of type A (type C) does R(11) of type B (type D). \\
\hspace*{5mm}As for R(10) and R(01), these spectra are nearly equal in the same type; it may be incidental.
Accordingly, from the same view point, R(10) and R(01) in four types are all the same.\\
\hspace*{5mm}The relationships between eight types of phase in four types and two types of unit cell  are also explained similarly.\\
\hspace*{5mm}The lattice constant of the models is 300 $\mu$m, and the frequency of the incident wave, for example, 0.4 THz is 750 $\mu$m in terms of wavelength. Two values are comparable rate.\\
\hspace*{5mm}When the wavelength of the incident wave is much larger than the lattice constant of a three-dimensional photonic crystal, however, the correlation of these reflecting features with two types of unit cell does not probably occur in general.\\
\section*{Conclusions}
\hspace*{5mm}The special characteristic features of reflected spectra and these phases on four types of cut-surface that were shifted per a/4 (a: lattice constant) in the same direction (001), were studied especially within photonic band gap by using the FEM. They were done per cut-surface type, especially with two types of orientation, [(11) and (1$\overline{1}$) in an XY-plane] of electric-field of the incident wave in normal incidence, and two types of unit cell which convert each other by 90-degrees rotation about the central axis parallel to the Z-axis were suggested.
As a result, it was found that eight types of reflected spectrum and those of phase in all four cut-surface types are able to be classified in two patterns, and they are correlated with above two types of unit cell. \\ 

\newpage
\section*{Appendix}
\setcounter{figure}{0}
\renewcommand{\figurename}{Fig.~A.}
\begin{figure}[h]
\centering
\includegraphics[width=6.5cm, trim=0 0 0 -100]{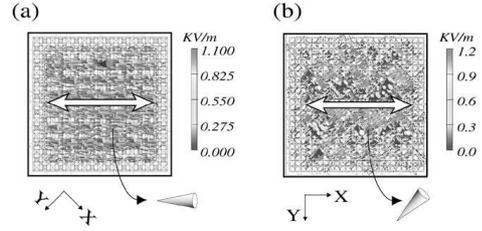}
\caption{\footnotesize Type A's electric-field vectors, cones on the bottom surface (Z=0) at 0.414THz. Input power is 1.0 (W). (a) model(1$\overline{1}$). (b) model(10).}
\label{fig:EEfield}
\end{figure}
\begin{figure}[h]
\centering
\includegraphics[width=7.0cm, trim=0 0 0 -150]{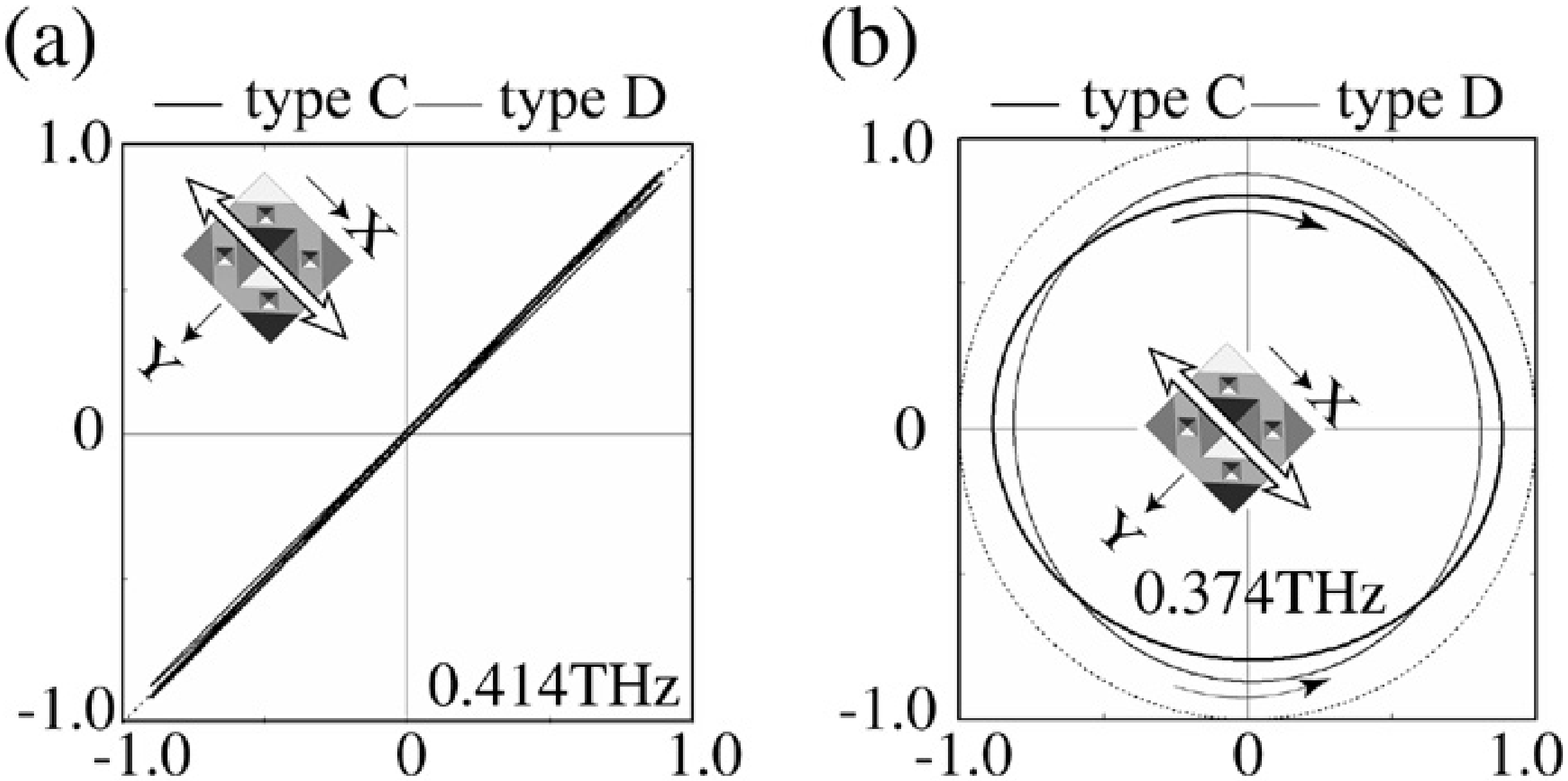}
\caption{\footnotesize Orbit of electric-field vector of the reflected wave. The white heavy arrow is the orientation of electric-field vector of the incident wave.
(a) at 0.414THz. (b)  at 0.374THz. The dashed line is true circle.}
\label{fig:typeCDorbit}
\end{figure}
\begin{figure}[t]
\centering
\includegraphics[width=6.5cm, trim=0 0 0 -30]{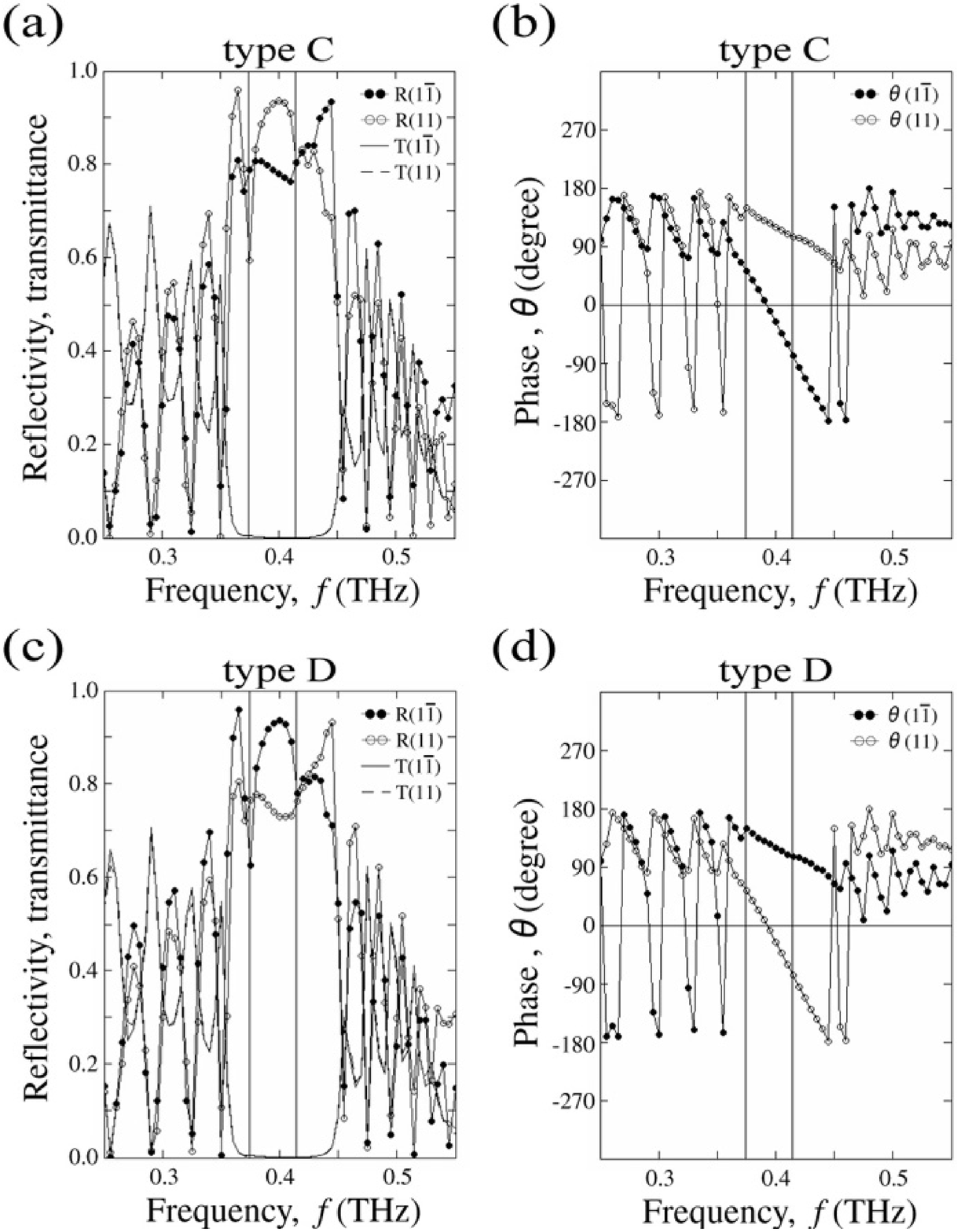}
\caption{\footnotesize Reflected spectra, R, those phases, $\theta$, and transmission spectra, T of type C and type D in model(1$\overline{1}$) and model (11) in fig.~\ref{fig:modelAB} (a).}
\label{fig:typeCD11}
\end{figure}
\begin{figure}[t]
\centering
\includegraphics[width=6.5cm, trim=0 0 0 0]{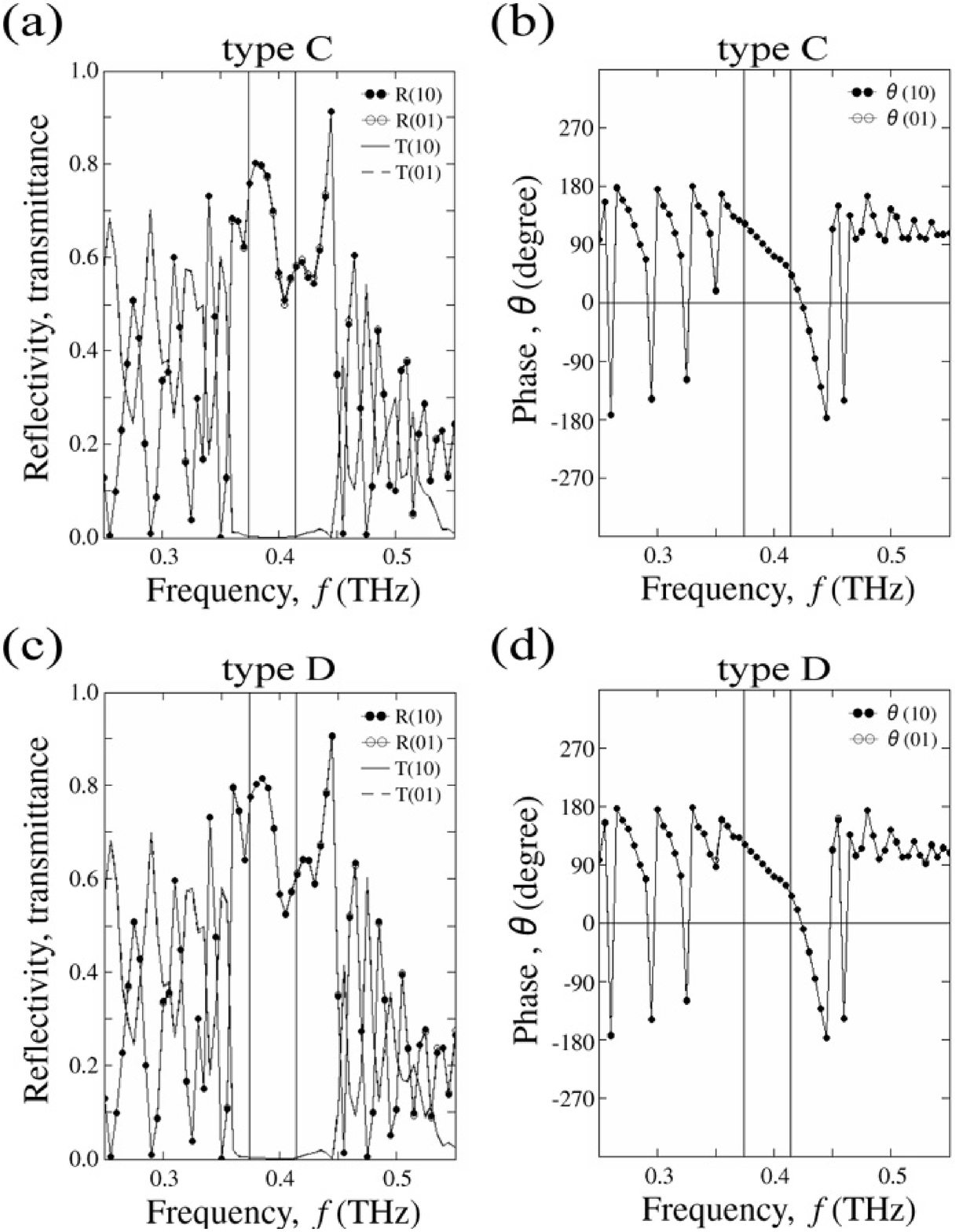}
\caption{\footnotesize Reflected spectra, R, those phases, $\theta$, and transmission spectra, T of type C and type D in model (10) and model (01) in fig.~\ref{fig:modelAB} (b).}
\label{fig:typeCD10}
\end{figure}
\newpage
\begin{figure}[h]
\centering
\includegraphics[width=6.5cm, trim=0 0 0 1500]{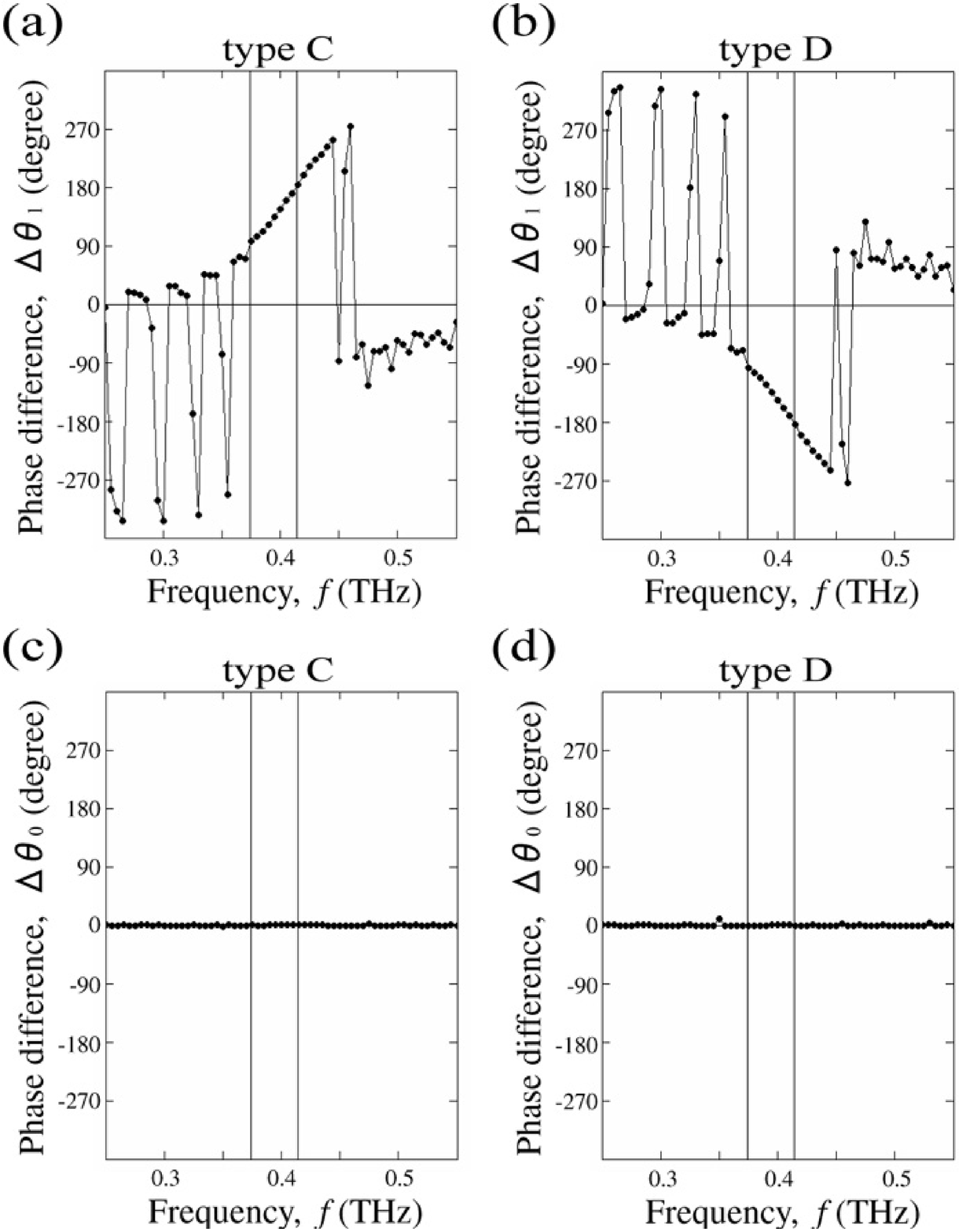}
\caption{\footnotesize Phase-differences of type C and type D. (a) $\Delta\theta_{1}$ is the phase-difference, $\theta(11)-\theta(1\overline{1})$.  (b) $\Delta\theta_{0}$ is $\theta(10)-\theta(01)$.}
\label{fig:typeCDd}
\end{figure}

\end{document}